\documentclass[draft]{agujournal2019}
\usepackage{url} 
\usepackage[finalnew]{trackchanges}
\usepackage{soul}
\usepackage{natbib}
\usepackage{siunitx}
\usepackage{subfig}
\draftfalse

\journalname{JGR: Planets}

\begin{document}

%
%


\title{On-deck seismology: Lessons from InSight for future planetary seismology}

%
%




\authors{M. P. Panning\affil{1}, W. T. Pike\affil{2}, P. Lognonn\'{e}\affil{3}, W. B. Banerdt\affil{1}, 
N. Murdoch\affil{4}, D. Banfield\affil{5}, C. Charalambous\affil{2}, S. Kedar\affil{1}, R. D. Lorenz\affil{6}, A. G. Marusiak\affil{7}, J. B. McClean\affil{2}, 
C. Nunn\affil{1}, S. C. St\"{a}hler\affil{8}, A. E. Stott\affil{2}, and T. Warren\affil{9}}


\affiliation{1}{Jet Propulsion Laboratory, California Institute of Technology}
\affiliation{2}{Imperial College, London}
\affiliation{3}{Universit\'e de Paris - Institut de Physique du Globe, CNRS, F-75005 Paris, France}
\affiliation{4}{ISAE-SUPAERO, Toulouse, France}
\affiliation{5}{Cornell University}
\affiliation{6}{Applied Physics Laboratory, Johns Hopkins University}
\affiliation{7}{University of Maryland}
\affiliation{8}{ETH Z\'{u}rich}
\affiliation{9}{Oxford University}




\correspondingauthor{Mark Panning}{Mark.P.Panning@jpl.nasa.gov}




\begin{keypoints}
\item Based on InSight recordings, atmospheric noise is amplified \change{when not placed on the ground}{for seismic sensors on deck}, consistent with Viking observations.
\item On Mars, this effect suggests long periods of observation (months to years) are required to detect \add{seismic} events.
\item On an airless body, like \remove{Jupiter's moon} Europa, on-deck or in-vault deployment \add{of seismic sensors} may be adequate to observe \add{seismic} events with days of observation.
\end{keypoints}

%
%

%
%


\begin{abstract}
Before deploying to the surface of Mars, the short-period (SP) seismometer of the InSight mission operated on deck for a total of 48 hours.  This dataset can be used to understand how deck-mounted seismometers can be used in future landed missions to Mars, Europa, and other planetary bodies.  While operating on deck, the SP seismometer showed signals comparable to the Viking-2 seismometer near 3 Hz where the sensitivity of the Viking instrument peaked.  Wind sensitivity showed similar patterns to the Viking instrument, although amplitudes on InSight were $\sim$80\% larger for a given wind velocity.  However, during the low wind evening hours the instrument noise levels at frequencies between 0.1 and 1 Hz were comparable to quiet stations on Earth, although deployment to the surface below the Wind and Thermal Shield lowered installation noise by roughly 40 dB in acceleration power.  With the observed noise levels and estimated seismicity rates for Mars, detection probability for quakes for a deck-mounted instrument are low enough that up to years of on-deck recordings may be necessary to observe an event.  Because the noise is dominated by wind acting on the lander, though, deck-mounted seismometers may be more practical for deployment on airless bodies, and it is important to evaluate the seismicity of the target body and the specific design of the lander.  Detection probabilities for operation on Europa reach over 99\% for some proposed seismicity models for a similar duration of operation if noise levels are comparable to low-wind time periods on Mars.
\end{abstract}

\section*{Plain Language Summary}
In the Viking-2 mission in the late 1970's, a seismometer was used on the deck of the lander but only saw one event that could be interpreted as a signal like earthquakes on Earth.  Because of this, the InSight mission put their seismic instrument on the ground and covered it up to keep the wind from blowing on it.  But we can use the time period where it was turned on before getting put on the ground to figure out whether future missions could do seismology without placing it on the ground.  We find that the wind blowing on InSight gave us similar signals to the Viking lander, even though InSight had a better instrument.  When we use models of how many quakes should be on Mars, we find that keeping the instrument on deck makes it hard to see any quakes unless we listen for months or years.  But we may be able to do better on planets and moons that do not have air and wind.  A lander on Jupiter's moon Europa, for example, could have a large chance of detecting events within a few days of recording even if the instrument is not put on the ground.

%
%

%


%
%
%
%

\section{Introduction}
The InSight mission to Mars landed on November 26, 2018 \citep{Banerdt+2020}. This geophysics mission was the first to deliver a seismometer \citep[SEIS,][]{Lognonne+2019} to the martian surface since the Viking landers in the 1970's \citep[e.g.][]{Anderson+1977,Nakamura+1979b,Lazarewicz+1981,Lorenz+2017b}.  While both Viking landers included a seismometer mounted on their deck, only one managed to uncage, and initially only one potential event with an internal origin was identified \citep{Anderson+1977}.  In the final study of the investigation following the mission \citep{Lazarewicz+1981}, the most critical shortcoming identified was the need to get the seismometer off the deck and directly coupled to the ground.

Prior to InSight, the importance of ground coupling was handled in different ways by projects which either failed after launch, like Mars 96, or were cancelled by the end of phase B, like NetLander. OPTIMISM \citep{Lognonne+1998b}, onboard the Autonomous Small Surface Mars 96 Stations \citep{Linkin+1998}, was mounted on the Small Station structure, expected to sit directly on the Martian surface. The small station had no feet like Viking, reducing the lack of rigidity proposed as the source of the Viking wind sensitivity \citep{Lognonne+1993}. The rigidity of the Small Station structure connecting OPTIMISM to the ground was however identified as critical for noise levels near $10^{-8}$ $\mathrm{m}/\mathrm{s}^2/\sqrt{\mathrm{Hz}}$. Consequently, a carbon fiber structure was therefore designed and integrated in the Small Station for better seismic coupling.

A precursor of SEIS \citep{Lognonne+2000} was considered for the proposed NetLander mission \citep{Harri+1999,Dehant+2004}. Although located inside the structure of the NetLander, the seismometer was designed with a lander mechanical decoupling device, enabling the seismometer to deploy three feet through holes in the floor of the lander which would penetrate the ground using the weight of the lander. The seismometer was then expected to be decoupled from the lander. Although much more risky than InSight, as no site selection could be made, and also less efficient in terms of lander noise reduction, this strategy was considered as optimum when no robotic arm was available. 

These examples illustrate three important parameters for the quality of a seismometer installation: (i)  the rigidity of the seismic path, between the sensor and the bedrock or surface,
(ii) the efficiency of the installation to attenuate the lander noise,
(iii) the efficiency of the installation to attenuate the seismic noise trapped in the low-velocity layer just beneath the surface.
With respect to (i), seismic deployments on Mars depended on the lander legs (Viking), carbon structure (OPTIMISM) and the SEIS feet (Netlander and InSight). See \citet{Fayon+2018} for a detailed model of the coupling properties of SEIS through its feet. With respect to (ii), Viking and OPTIMISM did not provide any lander noise attenuation, while the lander noise attenuation for the decoupled instruments depends on the distance between the feet of the seismometer and the locations where the lander is in contact with the ground: about 10 cm for Netlander and $\sim$1.5 meter for InSight. None of these installations attempted to mitigate (iii), which would require deploying the seismometer on bedrock or burying it.

In the context of all of these proposed seismic deployment methods, it should be emphasized that all of these approaches are reasonably well-coupled to the ground (point (i) above).  However on-deck deployments are also very well-coupled with the noise produced by the lander, both due to thermal effects and wind.  This is illustrated by the results of
an analog study instrumenting the engineering model of the Mars Science Laboratory rover \citep{Panning+2019}. 
With respect to the seismic path quality (i), results
demonstrated that on-deck seismometers can potentially accurately recover ground signals for frequencies below lander resonances. This could suggest that Viking may have detected more events with a better instrument and modern digital seismic waveform processing not possible for most of the returned data which was sent back in a compressed event format rather than full waveforms \citep{Lorenz+2017b}.  This is clearly observed during the night, when the amplitude of the Viking seismic data was close to the instrument resolution \citep{Anderson+1977}.
But critically and with respect to the lander noise attenuation and criteria (ii),
\citet{Panning+2019} showed that the on-deck recordings showed degraded coherence with the ground signal when slight ``wind'' due to air-conditioning occurred during daytime hours.

The robotic deployment arm of the InSight mission represents a complex engineering product, and the deployment process involved months of spacecraft operations, representing a significant cost-driver for the mission.  Given the seismological focus of the science goals of this mission, this was an important investment to make and has led to detection of many events \citep{Giardini+2020} \note{reference to Giardini et al. (2019) AGU abstract removed from here}. However, future landed missions to Mars and other planetary bodies will likely have other primary science goals, but could still land seismic instruments without investing in robotic deployment.  In this study we examine on-deck recordings from the InSight mission in the context of better characterizing our ability to use on-deck recordings both on Mars as well as other planetary bodies.  We can compare these measurements with previous on-deck recordings made in the Viking-2 mission.  We show that on-deck recordings on Mars would likely still have a hard time detecting events even with modern instrumentation and full waveform return due to wind noise, but on-deck deployments on seismically active airless bodies may be practical, although of course noise due to extreme temperature variation may be more important in these cases.

\section{Operation of the InSight SP seismometer on deck on Mars}
\begin{figure}
\noindent\includegraphics[width=0.7\textwidth]{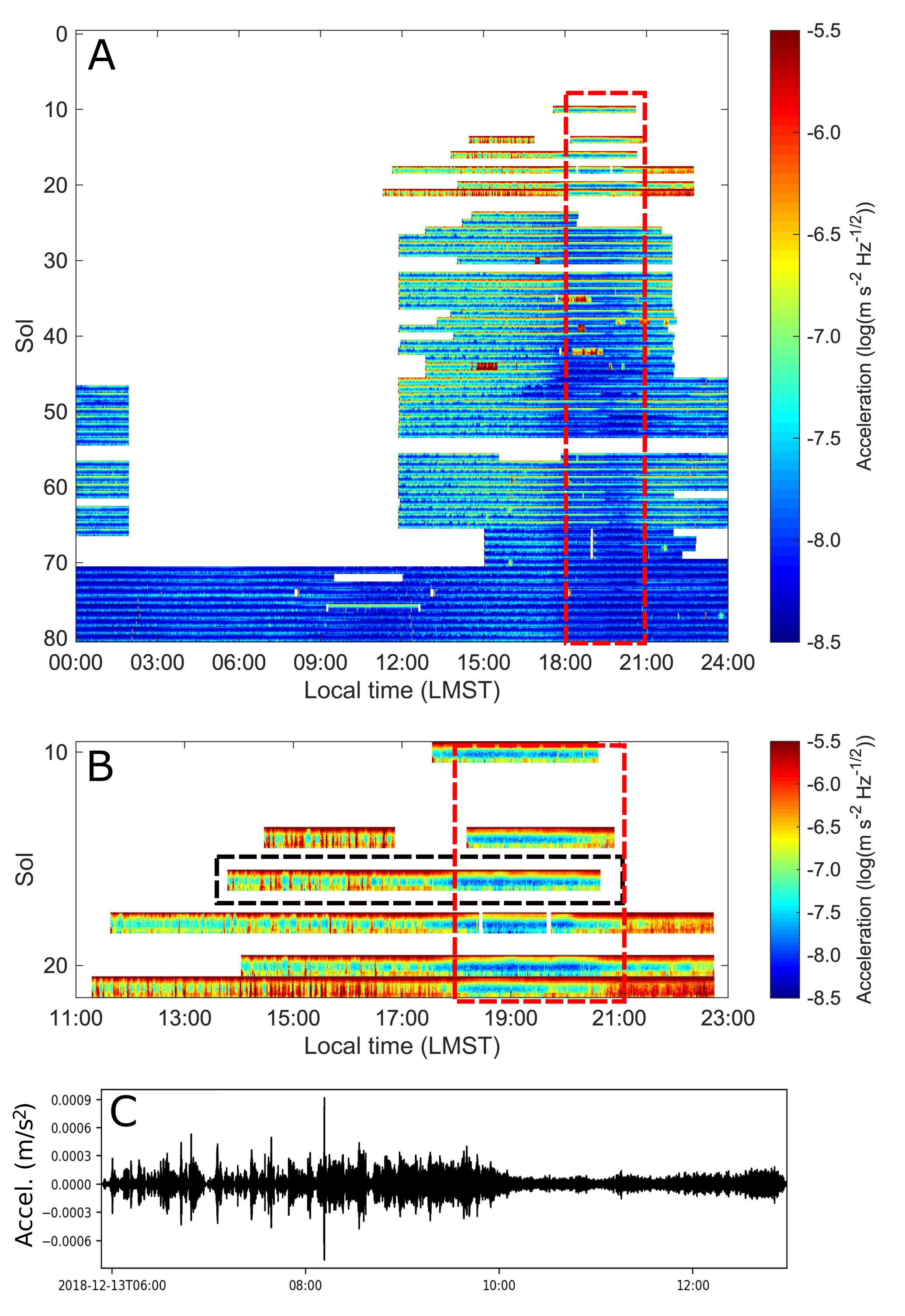}
\caption{\textbf{(A)} spectrogram of the acceleration as measured by SP1 (the vertical SP) on channel 68.SHU (corresponding to a sample rate of \SI{10}{\hertz}) versus Local Mean Solar Time (LMST), from sol 10 to 80. Seismometer ground deployment was on sol 22, and Wind and Thermal Shield (WTS) deployment was on sol 66. The first checkout on sol 4 at \SI{100}{\hertz} is not shown. \textbf{(B)} Detail of the spectrogram for data returned from the lander deck from sol 10 to 21. The red box in both panels indicates the evening quiet period observed consistently throughout the mission. \textbf{(C)} Ground acceleration seismogram filtered between 0.05 and 5 Hz for the sol in the dashed black box from panel B referenced to UTC time on Earth.}
\label{spectrograms}
\end{figure}

\begin{figure}
\noindent\includegraphics[width=0.8\textwidth]{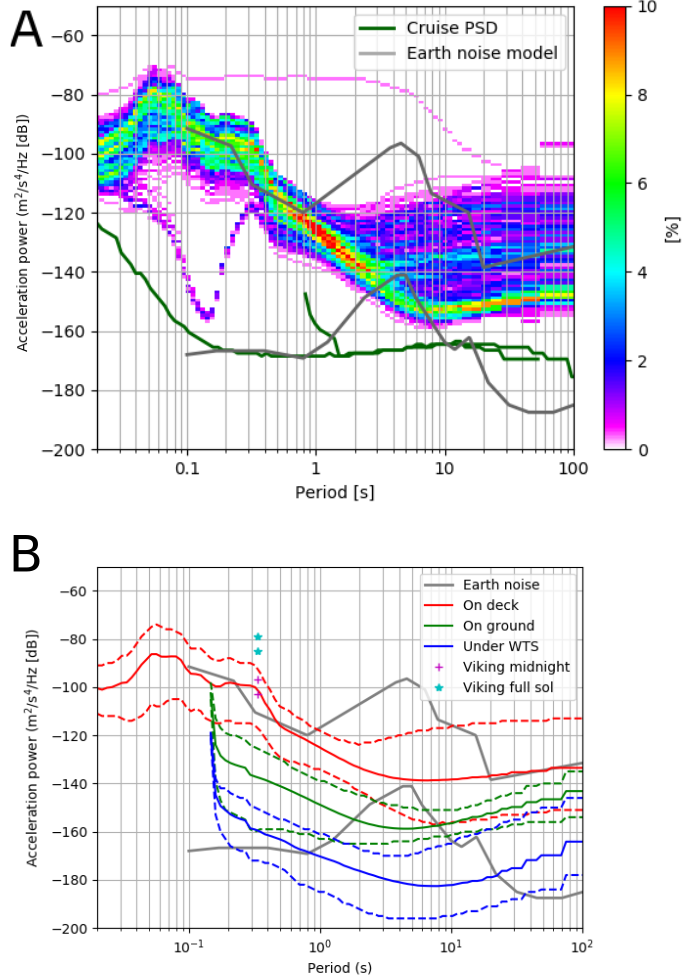}
\caption{\textbf{(A)} Probabilistic power spectral density \add{(PSD)} for the on-deck recordings of the InSight SP seismometer shown in color scale.  Note: outlier traces in light purple are related to calibration activities.  Grey lines show the low and high noise models for Earth data \citep{Peterson1993}, while green lines are the PSD for the SP recorded in cruise which represents the instrument self-noise. \textbf{(B)} Mean (solid lines) and 5\% and 95\% PSDs for the SP recorded on deck (red), the SP recorded on the ground prior to placement of the \change{WTS}{Wind and Thermal Shield (WTS)} (green) and 9 weeks of the VBB recorded under the WTS \add{(blue)} in February, March and April of 2019. Comparable amplitudes for Viking-2 records at the resonant frequency for that instrument  are shown in cyan and magenta (see section~\ref{viking} for the source of these numbers).  \add{All PSD estimates are calculated after deconvolution of instrument response.}}
\label{ppsd}
\end{figure}

Prior to being deployed on the surface, the InSight Short Period (SP) seismometer functioned on deck for almost 48 hours in the first $\sim$3 weeks of the mission after landing on Mars on November 26, 2018.  The Very BroadBand (VBB) instrument was not powered on while on deck due to leveling requirements in order to center the masses, while the SP is more tolerant of tilt \citep{Lognonne+2019}. Due to constraints on operating temperatures, the instrument could not be run continuously, but only ran during the daytime and early evening (Figure~\ref{spectrograms}).  While much of the day was typically quite windy, the red box in Figure~\ref{spectrograms}A and B highlights a time period that was consistently quiet on the seismic instrument on deck and on the ground before and after placement of the Wind and Thermal Shield (WTS). As shown in the continuous operated time periods after sol 70 in Figure~\ref{spectrograms}A, this evening quiet time on deck is consistent with the overall quietest time observed in the final deployment as well.  Overall, the noise observed on-deck was roughly 20 dB in power (a factor of 10 in amplitude) above the noise observed when deployed on the ground, while deployment of the WTS provided another 20 dB of noise reduction (Figure~\ref{ppsd}B).

One way of visualizing the noise level of seismic stations is to look at probabilistic stacking of power spectral density (PSD) of recorded signals over the duration in question \citep{McNamara+2004b}, as implemented in the python-based signal processing toolkit ObsPy \citep{Krischer+2015}. This is shown in Figure~\ref{ppsd}.  The signals are compared with Earth noise models in gray \citep{Peterson1993} and the mean power spectral density of the SP horizontal components when turned on during cruise to Mars.  When no trajectory adjustments were being made, the cruise recording was a quieter environment than is ever possible on Earth or Mars, and the observed noise matched pre-mission expectations of instrument performance, and so the PSD represents the self-noise of the instrument.  The vertical component was not tested in this way, as it requires Mars gravity for correct mass positioning.

The background noise recorded on deck at periods shorter than 1 second were generally comparable with high-quality Earth stations installed in noisy locations such as ocean islands, as represented by the New High Noise Model of \citet{Peterson1993} (upper gray line of Figure~\ref{ppsd}).  For periods longer than 2 or 3 seconds, however, noise levels varied much more widely between the windy afternoon time period and the quieter early evening.  In fact, noise levels between 3 and 10 second periods during the quiet period were frequently more than 10 dB quieter than the quietest stations on Earth as represented by the New Low Noise Model (NLNM) of \citet{Peterson1993}.  This is a remarkable observation given that this is comparing an instrument deployed on top of a meter-high lander deck exposed directly to the wind with the most carefully installed seismic vault and borehole sensors on Earth.  This emphasizes how much more seismically quiet Mars is than Earth in this frequency band, which is dominated by significant ocean wave noise on Earth called the microseism \citep[e.g.][]{Longuet-Higgins1950}.

The binning process used in the probabilistic PSD estimation smoothes over spectral peaks in the data which can be more clear in individual spectral estimates, but smoothed peak structure can be seen between periods of 0.03 and 0.3 seconds (i.e. $\sim$3-30 Hz in frequency).  This is related to lander modes discussed in section~\ref{sec_landermodes}.

\section{Comparison with Viking data}
\label{viking}
The two Viking landers, launched in 1976, both included seismometers, although the instrument on Viking-1 failed to uncage \citep{Anderson+1976,Anderson+1977,Lazarewicz+1981}.  The useful Viking-2 data, which was recently made fully available to the NASA Planetary Data System Geosciences Node \citep{Lorenz+2017b}, is primarily recorded in 2 modes, event and high-rate, with the vast majority being in the event mode.  The high-rate data was sampled at 20.2 Hz and includes the full waveforms. The event mode sampled at only 1.01 Hz, and actually returned the envelope of the amplitude signal at that sample rate along with the number of positive-going zero crossings, which allows the user to approximate amplitude and frequency content, but does not supply true digital waveforms.  It did, however, allow the instrument to send back amplitudes consistent with its maximum magnification near 3 Hz \citep{Anderson+1977}, as the event mode reduced data volumes by more than a factor of 10 compared to the high-rate mode.  The instrument's minimum resolvable ground motion was $\sim 2 \times 10^{-9}$ m in displacement ($\sim 7 \times 10^{-7}$ m/s$^2$ in acceleration) at 3 Hz, and $\sim 10^{-8}$ m or $\sim 4 \times 10^{-6}$ m/s$^2$ at 1 Hz \citep{Anderson+1976}.

\begin{figure}
\noindent\includegraphics[width=0.8\textwidth]{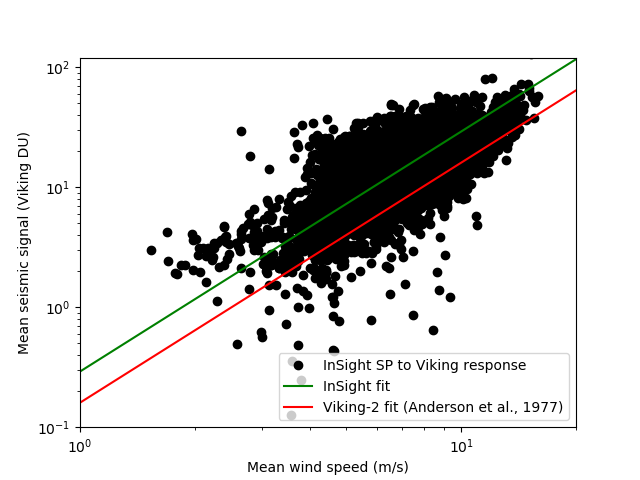}
\caption{Root mean squared wind speed as measured by the TWINS wind sensor compared with root mean squared seismic noise converted to the Viking instrument response.  Averages are made over 30 second windows for the time periods shown in figure~\ref{spectrograms}C with overlapping wind recordings (black circles).  The green line is the best linear fit with a fixed slope of 2, assuming seismic noise should scale with the squared wind velocity.  The red line shows the best fit to the original Viking-2 data taken from \citet{Anderson+1977}, figure 17.  See text for the equations for the InSight and Viking fits.}
\label{vikingwind}
\end{figure}

Because the Viking-2 instrument was located on the deck, it is useful to compare it with the signals we see from InSight on deck.  Overall we find a similar pattern on InSight as with Viking, with somewhat larger amplification of wind signals in the InSight data.  Diurnal patterns of amplitude appear to show somewhat similar patterns to InSight, with midday signals roughly an order of magnitude more noisy than during the night \citep[e.g.][fig. 4]{Lorenz+2017b}, which is comparable to the 20 dB range in power seen between the quietest 5\% and noisiest 5\% portion of the InSight data (fig.~\ref{ppsd}B).  While the Viking-2 seismometer operated for over 500 sols, the instrument only returned data from a few hours around midnight local time for much of that time \citep{Lorenz+2017b}.  For InSight, this time-period is typically much quieter than the midday time-period, but somewhat noisier than what is seen during the very quiet early evening hours \citep{Giardini+2020} \note{reference to Giardini et al. (2019) AGU abstract removed from here}. In the Viking data, for these limited sols, the peak amplitude observed is typically near 10-20 digital units (DU) \citep{Lorenz+2017b,Lorenz+2018}.  If we assume this is dominated by energy near the peak magnification of 3 Hz, this corresponds to ground displacement of $\sim 2-4 \times 10^{-8}$m or accelerations of $\sim 7-14 \times 10^{-6}$ m/s$^2$. For many sols in the first 60 sols and between sols 120 and 220, however, event mode for most of the diurnal cycle was returned, and peak amplitudes were closer to 60-125 DU, with the upper end controlled by the clipping of the instrument.  This would correspond to acceleration amplitudes of $\sim 5-10 \times 10^{-5}$ m/s$^2$ at 3 Hz.  These values are plotted at their equivalent power in Fig.~\ref{ppsd}B and are quite comparable to the on-deck recordings by InSight when considering that these are peak values rather than spectral power averaged over a finite time window, which would lead to higher values.  We can also compare our seismic recordings with the local winds as was done for Viking-2 \citep{Anderson+1977}.  Because the event mode data cannot be trivially converted to physical units, we instead convert the SP data on deck to the Viking instrument response (Viking DU) and compare with wind measurements from the InSight TWINS wind sensor (Fig.~\ref{vikingwind}).  In this comparison for both landers, we see a slope consistent with 2 on the log-log plot (corresponding to a correlation between seismic signals and squared wind velocity).  To make a quantitative comparison, the slopes of the two lines in Figure~\ref{vikingwind} are fixed at 2, and can then be expressed as 
\begin{equation}
y_{DU} = kU^2,
\end{equation}
where $y_{DU}$ is seismic amplitude in Viking digital units and $U$ is wind velocity in $\mathrm{m}/\mathrm{s}$, and $k$ is a proportionality factor, which is 0.16 DU/(m/s)$^2$ for the Viking-2 data and 0.29 DU/(m/s)$^2$ for InSight.  To approximate these proportionality factors in physical units, we can assume the Viking response data is dominated at the peak magnification at 3 Hz, and convert to acceleration using the 3 Hz magnification value ($2\times 10^{-9}$ m/DU in displacement or $\sim7.1\times 10^{-7}$ m/s$^2$/DU in acceleration) to obtain proportionalities of $1.1\times 10^{-7}$ (m/s$^2$)/(m/s)$^2$ for Viking and $2.1\times 10^{-7}$ (m/s$^2$)/(m/s)$^2$ for InSight.
This means that, for a given wind velocity, the seismic noise amplitude on deck for InSight is larger by a factor of $\sim$80\% than the equivalent Viking data.  This is consistent with the fact that the Viking lander had larger mass and no solar panels to catch the wind as InSight has.  

Overall, the signals are broadly consistent between the two landers.  Viking-2 is at a similar longitude to InSight, but at a much higher latitude and lower elevation, which suggests that the observed on-deck noise may not be strongly site-dependent, at least for sites chosen to be safe for landing. Steep slopes unfavorable to safe landing may be associated with strong katabatic winds, for example, and may have higher seismic noise levels as a result.

\section{Lander mode characterization}
\label{sec_landermodes}
The wind-induced mechanical noise has been recognised to be a potential problem for future space missions involving planetary seismometers, even when they are set on the ground \citep{Lorenz2012}. Long before the InSight mission, wind-induced noise was directly detected by the Viking seismic experiment on Mars \citep{Anderson+1977,Nakamura+1979b}. The Viking lander platform moved in the wind due to the low rigidity shock absorbers of the lander feet \citep{Lognonne+1993} and significant periods of time during the mission were dominated by the wind-induced lander vibration \citep{Goins+1979}. 

The wind-induced noise on the SEIS instrument of the InSight mission was studied in detail prior to landing \citep{Murdoch+2017a,Mimoun+2017,Murdoch+2018}. These analyses took into account the fact that the SEIS instrument would be positioned directly on the ground, and that the lander wind-induced noise would have to propagate elastically through the ground to SEIS. 

\begin{figure}
\noindent\includegraphics[width=0.8\textwidth]{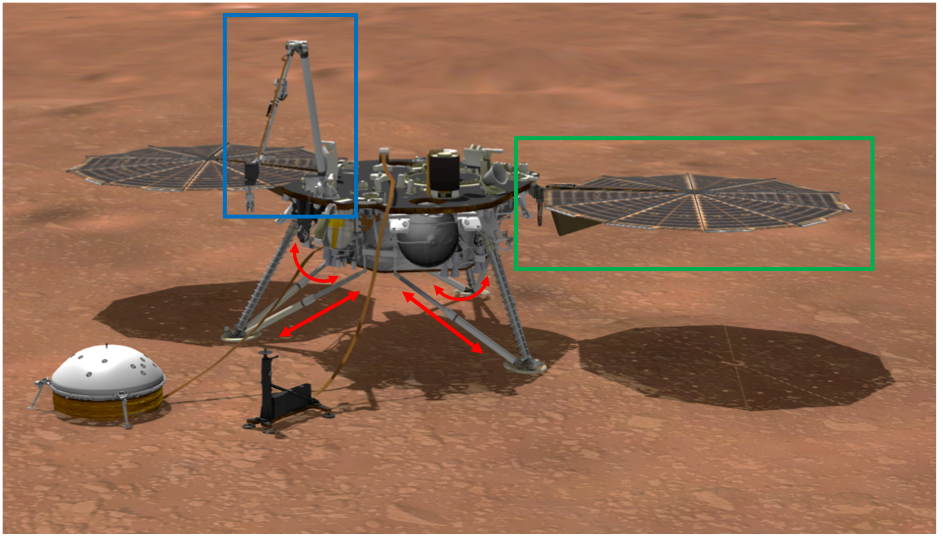}
\caption{An artist's impression of the InSight lander in its deployed configuration (courtesy: NASA). The solar panels and robotic arm, flexible appendages attached to the main structure, are shown by the green and blue boxes, respectively. The red arrows indicate the main directions of motion of the lander legs:  they can extend and contract along their length and the legs can splay apart and then come back together again.}
\label{lander-modes}
\end{figure}

One effect of the wind on Mars is the excitation of the resonant modes of the InSight lander. The InSight lander modes were required to be at frequencies above 1 Hz in order to be outside the very broad band seismometer bandwidth (0.01 - 1 Hz). However, many of these modes are visible in the short-period seismometer bandwidth (0.1 - 25 Hz). As part of the instrument commissioning, the SEIS short-period sensors were activated before the SEIS deployment onto the Martian surface. This provided a unique opportunity to observe the lander resonances while on the lander deck. However, it also highlighted the complex behavior of these modes and raised many questions as to their origins. 
There are multiple degrees of freedom in the InSight lander structure: the solar panels are flexible appendages attached to the main structure, the lander legs are flexible (for example, they can extend and contract along their length and the feet can slip over the ground as the legs separate and come back together), and the robotic arm is an additional flexible appendage (Fig.~\ref{lander-modes}). 

The frequencies of \remove{the} many of the resonances are linked to the temperature as the materials' rigidities vary as the temperature varies, 
and they are also linked to the ground properties: for the modes involving an interaction between the lander and the ground (a bounce-like motion, for example) a softer ground will lead to lower resonant frequencies whereas a stiffer ground results in higher resonant frequencies \citep[e.g.][]{Murdoch+2018}. The frictional properties of the foot - ground interface and the rotational coupling of the lander feet to the ground may also influence the frequencies of the resonant modes. 

Although a full numerical characterization of the resonant frequencies in a stowed and bolted configuration (i.e., the lander feet were assumed to be securely attached to the vibrating platform meaning that there was no degree of freedom between the feet and the vibrating surface) had been performed by Lockheed Martin to verify the launch environment requirements, no dedicated characterization tests were performed of the InSight lander in its deployed configuration.  After arrival on Mars, as the InSight seismometers recorded data both on the lander deck before being deployed to the ground and then in the deployed configuration on the ground, a study of the evolution of the modes was possible. This allowed a better identification of the modes originating from lander structure. However, given the large number of modes observed, and the difficulties associated with precisely identifying their origins, for any future on-deck seismology mission, it is highly recommended to fully characterize the resonant modes of the lander structure in a deployed configuration before launch. 

At the very least, this should be done by characterizing the structural response due to different input forces while the lander structure is fixed in place (i.e., bolted to a vibrating table). Given that the materials' rigidities are sensitive to temperature, ideally, these tests should also be performed in the operational temperature range of the lander. This will allow both a correct measurement of the frequencies of the modes, and possibly also some measurements of the evolution of these modes with the temperature variations. 
Finally, additional tests could also be performed on a `free' surface (unbolted lander) of different stiffness and or frictional properties to characterize the motion linked to the degrees of freedom between the feet and the ground.

\section{Probability of detecting events on Mars with only on-deck operation}
\label{marsprob}
In the time of observation on deck, no convincing observations of seismic events were made, consistent with the low or possibly zero detection rate of the Viking on-deck seismometer over a longer cumulative observation time \citep{Anderson+1977,Lazarewicz+1981}.  However, the Viking non-detection was also affected by the instrument quality, which was worse by roughly 2 orders of magnitude than the InSight SP at 3 Hz (the maximum magnification frequency of the Viking instrument) and 3-6 orders of magnitude in the quiet seismic band between 0.1 and 1 Hz \citep{NASA1976}, and the fact that most recordings were sent back in the compressed event format (see section~\ref{viking}).

Given the limitations of the Viking experiment, we would like to better evaluate the probability of detecting a seismic event with the SP recording on deck for the actual observation time period, as well as assessing the likelihood of observation over longer time-windows possibly accessible to future long-lived landers without a mechanism for deploying a seismic package directly on the surface.

While events have subsequently been detected by SEIS after deployment to the surface and placement of the WTS 
\citep[e.g.][]{Lognonne+2020,Giardini+2020} \note{References to AGU abstracts Lognonne et al. (2019) and Giardini et al. (2019) were removed from here},
allowing for initial estimations of Mars' seismicity, there are still only a small number of Marsquakes observed.  This means that magnitude estimates remain relatively uncertain, and well-calibrated attenuation curves to estimate amplitude of signals as a function of distance from the source are still not available.  In fact, initial estimates of quake amplitudes are based on scales calibrated using synthetic Mars seismic data from prior to InSight landing \citep{Bose+2018}.  Given this current limitation, we choose to initially also use attenuation curves derived from synthetic data in order to further explore the potential for on-deck seismology from a statistical perspective.

\begin{figure}
\noindent\includegraphics[width=0.8\textwidth]{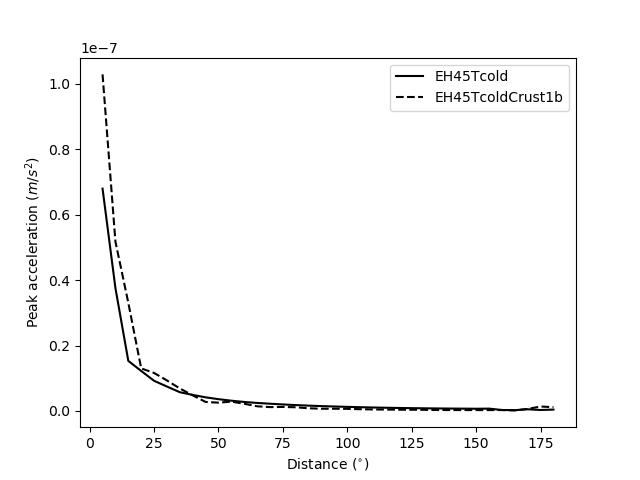}
\caption{Attenuation curves showing peak acceleration amplitude as a function of distance for 2 Mars interior structure models with Instaseis databases computed for the Marsquake Service blind test \citep{Rivoldini+2011,Clinton+2017}.  The two models have very different crustal structure, which generally creates the largest contrast in predicted amplitudes over the range of models considered.  Amplitudes are calculated for a seismic moment of $10^{13}$ Nm, which can be linearly scaled to other event sizes.  Remaining figures in the paper use the curve for the \textit{EH45Tcold} model.}
\label{atten_curve}
\end{figure}

We develop an initial estimate of an attenuation curve based on pre-mission Mars interior models simulated using Instaseis \citep{vanDriel+2015}, which is a package that takes seismic waveform databases generated by the 2D numerical wave propagation code AxiSEM \citep{NissenMeyer+2014} and rapidly generates synthetics for arbitrary source mechanisms and source and receiver locations.  An attenuation curve is created by averaging amplitudes of signals for a given seismic moment over a range of randomly generated seismic faulting sources (Figure~\ref{atten_curve}).  Amplitudes then scale linearly as a function of actual seismic moment, so such an attenuation curve can then be used to estimate observed amplitude as a function of event moment, $M_0$, and epicentral distance, $\Delta$.  For the calculation shown in Figure~\ref{atten_curve}, the model EH45Tcold \citep{Rivoldini+2011,Clinton+2017,Smrekar+2019} was used, but most 1D models produce amplitudes that vary by a factor of 2 or less and have only a small impact on the results.  A much larger source of error actually arises from the surface wave amplitude.  1D synthetics from a range of a priori models \citep[e.g.][]{Smrekar+2019} are generally dominated between 0.1 and 1 Hz by unrealistically large surface waves that would not be expected on a planet with realistic 3D structure.  Indeed, early events observed by InSight have not yet seen detectable surface waves \citep{Lognonne+2020, Giardini+2020} \note{References to AGU abstracts Lognonne et al. (2019) and Giardini et al. (2019) were removed from here}.  For this reason, we choose to reduce the predicted amplitudes by a factor of 10 as first guess of a more realistic attenuation curve.  Regardless, the amplitude of this curve remains a significant uncertainty in the following work, likely meaning uncertainty in amplitudes from a factor of a few up to an order of magnitude. 
This estimated amplitude can be compared with noise amplitudes as determined by the mean value of the PSD estimates for each time segment of the data recorded on deck. When the amplitude (as predicted by synthetics) is compared with the noise according to some criteria, we can estimate the maximum distance  that a given amplitude could be recorded and therefore the fraction of the surface area of the planet that we could see.   

In order to estimate the detection probability as a function of seismicity, we first need to define our seismicity estimate.  There are many ways to do this, but in general, seismic events in a catalog usually follow a power-law distribution \citep[e.g.][]{Golombek+1992, Knapmeyer+2006}, $N(M_0) = AM_0^{-B}$, where $N(M_0)$ is the number of events greater than or equal to seismic moment $M_0$, and $A$ and $B$ are empirically determined coefficients for each seismicity catalog.  On Earth, we frequently express this as the Gutenberg-Richter relationship \citep{Gutenberg+1944},
\begin{equation}
\label{gr}
\log N(M_W) = a - bM_W
\end{equation}
where $M_W$ is the moment magnitude, defined by \citet{Kanamori1977} as
\begin{equation}
\label{momentmag}
\log M_0 = 1.5M_W + 9.1,
\end{equation}
where the seismic moment, $M_0$ is expressed in units of Nm, and $a$ and $b$ are empirically defined coefficients.  On Earth, $b$ values usually range between $\sim 0.7$ and $\sim 1.3$ \citep{Frohlich+1993}, although higher and lower $b$ values may be possible in many planetary settings (see \citet{Panning+2018} for further discussion).  With $b$ values near 1, there are generally a factor of 10 fewer events for each unit increase in magnitude.  Because of the factor of 1.5 in the definition of moment magnitude (equation~\ref{momentmag}), though, the energy increases by a factor of more than 30 for each unit increase of magnitude.  This means that setting the maximum magnitude of a catalog is needed to calculate a mean moment release rate for a given catalog as the large, rare events will dominate the total energy release \citep[e.g.][]{Golombek+1992}.  However, if we are only interested in the probability of observing events, this is less important than the traditional $a$ and $b$ values, as the large events will have vanishingly small probability of occurrence.

For a given seismicity model of Mars, defined by $a$ and $b$ values in equation~\ref{gr}, we can determine the probability of detecting $k$ number of events by assuming earthquakes are a Poisson process \citep{Poisson1837}, which is \change{a generally}{generally a} good assumption for Earth catalogs after removing aftershocks \citep[e.g.][]{Gardner+1974}.  In that case the probability of observing $k$ earthquakes in a given time period where we have an expected number of observations $\lambda$ is
\begin{equation}
\label{poisson}
P(k) = e^{-\lambda \frac{\lambda^k}{k!}}.
\end{equation}
In this case, where we are interested in the probability of observing at least 1 event, the relevant probability is the cumulative value for all $k$ greater than 0, which is simply $1-P(0)$.

In order to estimate $\lambda$ in equation~\ref{poisson} for a given value of $a$ and $b$, we need to estimate the expected number of detected events as a function of $M_W$.  We can estimate $\lambda$ for a series of magnitude bins of width $\Delta M$ for a given set of Gutenberg-Richter parameters as $\lambda(M_W) = N(M_W) - N(M_W + \Delta M)$.
Above some threshold magnitude, we should see all events on Mars regardless of the location, but for smaller events, we will not see events that are too far away.  In order to account for this, we define an amplitude detection threshold, $A_d = \gamma A_n$, where $A_n$ is the amplitude of noise estimated from the mean of the PSD for each window of data (Figure~\ref{ppsd}), and $\gamma$ is a minimum signal to noise ratio (SNR).  In this case, we choose $\gamma = 5$, which is rather high compared to our actual observation of Marsquakes, where observations are routinely made with SNR values less than 2 \citep{Giardini+2020} \note{reference to Giardini et al. (2019) AGU abdstract removed here}, but chosen because we're estimating amplitude with peak acceleration of 1D synthetics and comparing with mean noise amplitude, not peak noise amplitude.  Although we have already reduced these synthetic amplitudes to account for overestimation of short-period surface waves in 1D synthetics, the choice of this threshold remains a large source of uncertainty.  We can determine the maximum detection distance, $\Delta_d$ as a function of $M_0$, by setting $A_d = A(\Delta_d, M_0)$ in the attenuation curve (Figure~\ref{atten_curve}).  From that, we reduce the expected $\lambda$ for that magnitude bin by the fraction of surface area of the sphere covered by the maximum distance, 
\begin{equation}
\lambda_{\mathrm{eff}}(M_W) = \frac{1}{2}(1 - \cos \Delta_d)\lambda(M_W).
\end{equation}
The final value of $\lambda$ then for each choice of Gutenberg-Richter parameters is then simply the summation of $\lambda_{\mathrm{eff}}$ over all possible magnitude bins.  For the synthetic signals for a range of models like those used in the pre-landing Marsquake Service blind test \citep{Clinton+2017}, expected number of event detections drops off rapidly below $M_W=2$ due to small $\Delta_d$, and above magnitude 4-5 due to the drop-off in the Gutenberg-RIchter relationship, so $\lambda$ is dominated by magnitude bins between those limits.  For completeness, we consider all bins between $M_W = 0$, and $M_W = 8$, but the signal is dominated by events between 2 and 5.5.

This estimate relies on 3 factors: (1) the noise of the instrument, including that due to the motion of the lander, (2) the activity level of the planet, which is still not known for Mars, but is better constrained than before the mission \citep{Giardini+2020} \note{reference to Giardini et al. (2019) AGU abstract removed from here}, and (3) the attenuation and propagation characteristics of Mars.  We also assume Poisson statistics and have significant uncertainty in estimating detection threshold, primarily due to uncertainty in estimating signal amplitude due to propagation and attenuation effects.  Given all of these limitations, estimations of detection probability should be considered only as order of magnitude estimates, but they are based on real observed noise from the InSight mission.

\begin{figure}
\noindent\includegraphics[width=0.8\textwidth]{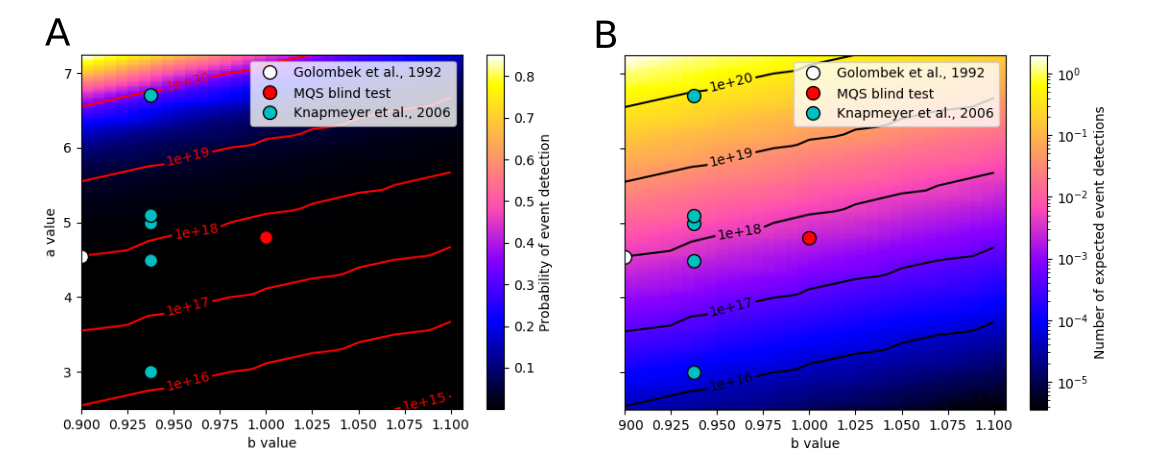}
\caption{Detection probability \textbf{(A)} and expected number of events detected  \textbf{(B)} \add{during the 48 hours of InSight on-deck observation} as a function of the assumed $a$ and $b$ parameters in the Gutenberg-Richter seismicity estimate (equation~\ref{gr}).  For comparison, predicted Mars seismicity rates from \citet{Golombek+1992} (white), \citet{Knapmeyer+2006} (cyan), and the seismicity rate in the Marsquake Service blind test \citep{Clinton+2017} (red) are shown.  Probabilities are calculated assuming a Poissonian process with detection thresholds defined as described in the text. \add{Contours (red in (A) and black in (B)) show the cumulative global seismic moment release per Earth year.}}
\label{detection}
\end{figure}

Figure~\ref{detection} shows our estimated probability of detecting at least one event (panel A) and the expected number of event detections for range of $a$ and $b$ values in the Gutenberg-Richter relationship.  Of the pre-event Mars seismicity estimates, only the highest estimate of seismicity from \citet{Knapmeyer+2006} shows a detection probability greater than 5\% (Figure~\ref{detection}A), and our actual observations of Mars events have demonstrated this model has unrealistically high levels of seismic activity \citep{Giardini+2020} \note{reference to Giardini et al, (2019) AGU abstract removed from here}.  For more realistic estimates of seismicity, our expected number of observations in the on-deck observation time are likely less than 0.001 (Figure~\ref{detection}B).  This suggests that on-deck observation would likely need to be continued for multiple Earth or even Mars years in order to have a significant probability of event detection.  This indicates that the deployment approach of InSight was extremely valuable as it dropped the noise levels by $\sim$40 dB in power, which corresponds to $\sim$2 orders of magnitude in amplitude, and made observations of seismicity on Mars possible.  However, this does not mean on-deck deployment of seismometers on other landed missions to other planetary bodies would not be useful. In particular, this dataset shows that the lander noise generated by wind is the critical factor in noise on the seismometer, and only Venus and Saturn's moon Titan are likely targets for future seismic deployments which have significant atmospheres.  Airless bodies may be attractive targets for on-deck deployment of seismic instruments.

\section{Implications for planetary seismology on airless bodies}
As demonstrated in Figure~\ref{vikingwind} and discussed in \citet{Anderson+1977}, the noise observed on deck for both InSight and Viking is dominated by the wind.  Given the seismicity level of Mars, this indicates that quake detection from a deck deployment is low in the absence of a deployment covering hundreds of days.  However, as demonstrated in \citet{Panning+2019}, deck-mounted seismometers record ground motions very well in the absence of wind.  This could imply that a deck-mounted seismometer could be quite useful on airless planetary bodies, like the Earth's moon, the icy moons of Jupiter, or Enceladus, a moon of Saturn.  Event detection probability would depend on the seismicity of the target body and its seismic propagation properties.  Landed seismic stations could also be particularly interesting for probing the internal structure of small-bodies. Indeed, several such concepts have been proposed in recent years using either explosive devices to generate seismic signals \citep[e.g.][]{Robert+2010}, or simply relying on the anticipated natural seismic activity of asteroids \citep[e.g.][]{Murdoch+2017c}.  The airless surface environment will also contribute to reducing the ambient noise on these targets, although the biggest challenges for such small-body seismic stations are likely to be the ground coupling in an extremely low-gravity environment and the often extreme temperature variations \citep{Cadu+2016}.

As an example of this, \change{we'll}{we} look in detail at potential detection probabilities for a deck-mounted instrument on Europa.  \citet{Panning+2018} estimated a range of seismicity estimates for Europa, based on an assumed scaling based on estimated tidal dissipation energy compared with the observed seismicity on the Earth's moon \citep[e.g.][]{Oberst1987}.  Attenuation curves can be calculated in the same fashion as in section~\ref{marsprob}.  For the structure model, we use a 20 km thick ice shell Europa model from \citet{Vance+2018b} and an Instaseis database from \citet{Stahler+2018}, and make the same assumption for reduced surface wave amplitude due to 3D structure that was made for the Mars data.  Obviously, there is significant uncertainty in this estimate, but this uncertainty is likely much smaller than the uncertainty in seismicity estimates, which span 2-3 orders of magnitude.

\begin{figure}
\noindent\includegraphics[width=0.8\textwidth]{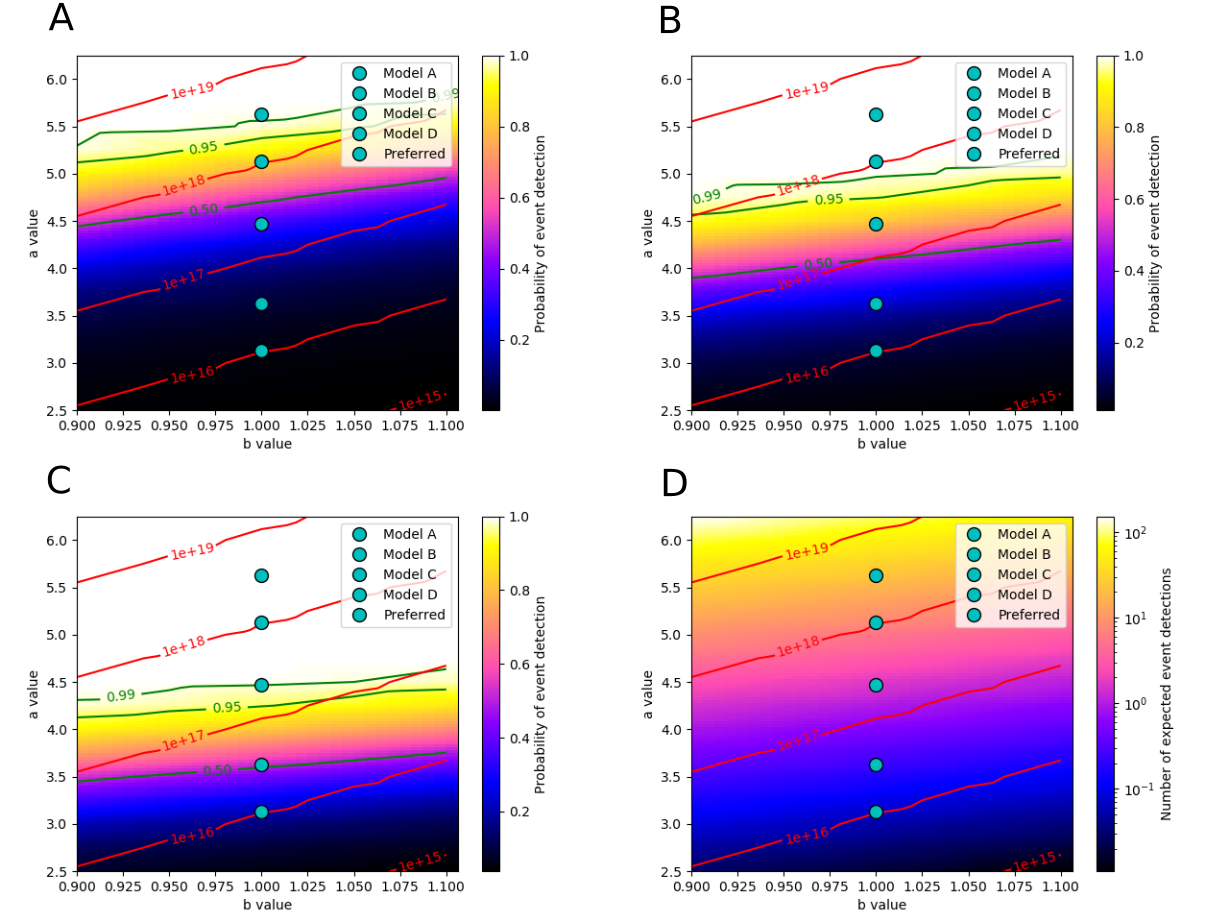}
\caption{Event detection probability on Europa for 48 hours of data calculated in a similar manner to the Mars estimates in Figure~\ref{detection}.  The synthetics for attenuation curve calculation are from an Instaseis database based on a Europa structure model from \citet{Vance+2018b} as used in \citet{Panning+2018}.  Instrument noise is assumed to match that actually observed on-deck on  Mars \textbf{(A)}, continuously remain at the lowest observed noise on Mars \textbf{(B)}, or match the self-noise of the InSight SP instrument \textbf{(C)}.  Expected number of event detections for the low noise model of panel (B) is shown in panel \textbf{(D)}.  Cyan dots show a range of predicted Europa seismicity models from \cite{Panning+2018}.  Red lines show contours of cumulative global moment release per Earth year, while green lines show the 50\%, 95\% and 99\% probability contours.}
\label{Europa_detection}
\end{figure}

In order to compare with event detection probabilities for Mars, we assume the same total observation time as the InSight on-deck recordings (48 hours).  If we simply assume the same instrument noise as on Mars (Fig.~\ref{Europa_detection}A), detection probabilities in 48 hours are as high as 99\% for the highest seismicity model, while the preferred seismicity model of \citet{Panning+2018} would suggest a detection probability of just under 50\%.  A better guess for instrument noise may be to use the low-noise portions of the on-deck recordings. The high-noise portions are dominated by wind, which would not be present on an airless body. The remaining noise may be more influenced by thermal noise within the lander itself, and is also expected for airless bodies.  In that case (Fig.~\ref{Europa_detection}B), the detection probability increases to over 80\% for the preferred seismicity model, while a bounding scenario would be noise matching the self-noise of the SP seismometer (Fig.~\ref{Europa_detection}C).  In this case, even the most pessimisitic seismicity model shows a detection probability of a few percent, while the preferred seismicity model shows a 99\% detection probability in 48 hours.  A proposed mission to Europa, the Europa Lander \citep{Hand+2017}, would include a seismic instrument in its baseline and threshold missions and would have a proposed duration of surface operations of a few weeks.  Assuming the seismometer is recording during most of that duration, it would be reasonable to expect 5-10 times longer total observation time than considered in Figure~\ref{Europa_detection}.  This strongly suggests that a deck-mounted seismometer would have a high probability of recording quakes in a similar landed mission to Europa.

\section{Discussion}
At the time of writing (2019) development is starting on a new mission planning to make seismic observations. Dragonfly \citep{Lorenz+2017,Turtle+2018} would be a lander which explores the habitability and prebiotic chemistry of Titan. Of interest in this context is the thickness of Titan's ice crust, overlying an internal water ocean. Dragonfly is a rotorcraft, and although its concept of operations with relocation flights every Titan day or two (i.e. approximately once per month) precludes an elaborate instrument emplacement, it is planned to have a seismometer lowered to the ground with a windshield by a belly-mounted winch.  In addition, geophones are mounted on its landing skids.  While these will be affected by wind loads on the lander, they will at least be well-coupled to the ground. Furthermore, the lander noise signals measured by the geophones directly can be used to estimate, and decorrelate, the ground-coupled lander noise sensed by the seismometer. There may be extended periods where seismometer observations are made before or after winch activation.  On these occasions, the seismometer will be functioning in an `on-deck' mode. Given the presence of a dense atmosphere on Titan, the issue of lander noise clearly deserves further study.  In particular, detectability will depend upon the expected seismicity of Titan, which is preliminarily estimated (based on tidal dissipation energy) to be lower than Europa \citep{Hurford+2020} and the wind noise.  Windspeeds near the surface of Titan are directly constrained by data from the trajectory of the Huygens lander \citep[e.g.][]{Karkoschka2016}, and are also modeled through global circulation models \citep[e.g.][]{McDonald+2016, Lora+2019}, and are expected to be of the order of 0.1 to 1 m/s.  If windspeeds are 10-100 times slower on Titan, but the atmosphere is denser by a factor of a few hundred, dynamic pressures (proportional to density times wind speed squared) may be of a similar order of magnitude on Titan and Mars, but clearly more detailed modeling is needed.

In this study, we have chosen to focus on the ability to detect internal events, but a seismic instrument can also use other signals to perform useful science.  Correspondence between pressure and seismic signals (e.g. \citet{Sorrells1971} on Earth or \citet{Lognonne+2020} \note{Reference to Lognonn\'{e} et al. (2019) AGU abstract removed here} for Mars) can reveal subsurface elastic properties.  Observation of processes that interact with the surface, like InSight's mole \citep{Spohn+2018}, can be used like an active source survey can also be used to constrain the near surface structure.  Science goals for deck-mounted instruments need to be carefully considered to take into account the wind-driven noise.

The detection probabilities estimated here in detail for Mars and Europa assume uniform distribution of seismicity.  This is obviously not true for the Earth, where seismicity is focused on plate boundaries.  It's probably not true on Mars, where observed surface faulting and ages is heterogenous \citep[e.g.][]{Knapmeyer+2006}.  If icequakes on Europa are driven by tidal cracking, it's reasonable to think seismicity may follow the distribution of tidal energy \citep{Hurford+2020}.  The non-uniform distribution would mean that detection probability would depend on landing site location.  For the tidal dissipation modeling in \citet{Hurford+2020}, this only implies variations on the order of 15\%, but the localization observed on Earth is much larger effect, and so the estimates in this study would only represent global averages.

\section{Conclusions}
Below the resonant frequencies of the lander, on-deck or in-vault seismometers accurately record ground motion.
Atmospheric noise (particularly wind) is amplified when not placed on the ground, as well as other lander activity noise.
This recording of lander noise by InSight is consistent with the wind signals recorded by the Viking-2 seismometer, with a similar dependence on the square of wind speed at slightly higher amplitude in the frequency band of Viking sensitivity. 
On Mars, this effect is important enough that we would be unlikely to record any events without recording continuously for one or more years, suggesting the lessons of Viking were not too far off, and the deployment strategy of InSight was important to observe seismic activity.
On an airless body like Europa, though, deployment on the ground seems less necessary (with the caveat that the lander may generate lots of internal noise).

\acknowledgments
MPP, WBB, SK, and CN were supported the NASA InSight mission and funds from the Jet Propulsion Laboratory, California Institute of Technology, under a contract with the National Aeronautics and Space Administration. WTP, CC, JBM and AES were supported by the UK Space Agency. The French authors acknowledge CNES and ANR(\change{MAGIS}{ANR-14-CE36-0012-02 and ANR-19-CE31-0008-08}) for their support. We acknowledge NASA, CNES, partner agencies and Institutions (UKSA, SSO, DLR ; JPL, IPGP-CNRS, ETHZ, IC, MPS-MPG) and the operators of JPL, SISMOC, MSDS, IRIS-DMC and PDS for providing SEED SEIS data. InSight seismic data presented here (\url{http://dx.doi.org/10.18715/SEIS.INSIGHT.XB_2016}) is publicly available through the Planetary Data System (PDS) Geosciences node, the Incorporated Research Institutions for Seismology (IRIS) Data Management Center under network code XB and through the Data center of Institut de Physique du Globe, Paris (http://seis-insight.eu).  Example scripts for processing data are included in the GitHub repository https://github.com/mpanning/OnDeckInSight (archived version for this submission is at DOI: \url{https://doi.org/10.5281/zenodo.3712408}, and uses data archived at DOI \url{https://doi.org/10.25966/na0g-cr11}). This is InSight Contribution Number 120.  Copyright \change{2019}{2020}. All rights reserved.


%
%

\bibliography{biblio}

%
%
%
%
%

\end{document}